\documentclass[12pt,a4paper]{article}
\usepackage{amsmath}
\usepackage{graphicx}
\usepackage{times}
\begin{document}
\begin{titlepage}
\title{Reflecting disk with black edge}
\author{ S.M. Troshin, N.E. Tyurin\\[1ex]
\small  \it SRC IHEP of NRC ``Kurchatov Institute''\\
\small  \it Protvino, 142281, Russian Federation}
\normalsize
\date{}
\maketitle

\begin{abstract}
We  discuss geometrical picture of hadron interactions and, in particular, the presence of energy--invariant edge in elastic scattering resulting from a transition to the reflective scattering mode which is supported by  the recent LHC data. The energy-invariant edge represents  a black ring around the reflecting disk.
\end{abstract}
\end{titlepage}
\setcounter{page}{2}
\section*{Introduction}
The geometrical approach to the studies of soft hadron interactions and the theory of elastic processes based on geometrical concepts have been quite successful and allowed one to explain many of the experimental results since the seminal papers by Chou and Yang  have been published 
\cite{y1,y2,y3}.
The recent analysis of the elastic scattering performed in the impact parameter representation has led to consideration of the 
``edge'' in elastic scattering. This  has been  discussed in \cite{b1,b2,ros,fag}. It was argued that  ``soft edge'' has an energy-invariant form and gives a subdominant contribution to the rise of the total cross--section. It has also been argued that the asymptotical picture of the proton
scattering corresponds to the black disk with the soft edge. Energy independence of this edge is consistent with BEL (Blacker, Edgier, Larger) picture  \cite{valin,valin1} since the {\it relative} contribution of the edge is decreasing with the collision energy. In case of the black-disk picture, the edge
can be associated \cite{b2} with the functional dependence establishing the   upper bound on the inelastic diffractive cross--section.

As it was noted, the elastic scattering  data are consistent with the above BEL picture when the protons become blacker, relatively edgier and larger.   The analysis of the data on elastic scattering obtained by the TOTEM  at $\sqrt{s}=7$ TeV has pointed out to an existence  of the new regime in strong interaction dynamics,  related to transition  at very high energies to the new scattering mode described in  \cite{ph1,phl,degr,intja} and referred to as
antishadowing or reflective scattering. It corresponds to the gradual transition to the REL picture, i.e. when the interaction region starts to become reflective at the center and simultaneously relatively edgier, larger and black at its periphery. Experimentally, its appearance is manifested under a reconstruction of the elastic amplitude, elastic and inelastic
overlap functions in the impact parameter representation  \cite{alkin}.  

This is the most sensitive method since it is based on the analysis of the {\it differential} cross--sections. Another relevant quantities are the ratios of the elastic cross-section and slope parameter
$B(s)$ to the total cross--section $\sigma_{tot}(s)$.  The ratio $\sigma_{el}(s)/\sigma_{tot}(s)\to 1/2$ 
and
${16\pi B(s)}/{\sigma_{tot}(s)}\to 1/2$
at $s\to\infty$ in case of black--disk limit saturation, while the limiting values of the both  ratios are unity  in case of
reflective scattering domination at $s\to\infty$. Actually, the experimental values for the former ratio are increasing while the values of the latter quantity are decreasing with the collision energy (in the region of high energies) (cf. e.g. \cite{dremin}). However, any studies of these quantities (since the both  ratios include integrated over the impact parameter 
functions) are less sensitive tools  for detection of  gradually emerging  deviation from the black- disk limit which initially occurs in the narrow region of the impact parameter variation.
It should also be noted that the recent
luminosity-independent measurements performed at $\sqrt{s}= 8$ TeV \cite{8tev} confirmed a steady increase of the ratio $\sigma_{el}(s)/\sigma_{tot}(s)$. This increase is another but, of course,  indirect indication on existence of the reflective scattering mode. 
However, the analysis of the TOTEM data for the {\it differential} cross--section \cite{alkin}  provides a rather strong indication to the  asymptotics associated with the reflective scattering mode. 

In view of  the above said, we discuss in this note  the edge properties in the elastic scattering in the case when the reflective scattering mode being present. 
\section{The edge in  absorptive and reflective scattering}
The elastic scattering $2\to 2$ $S$-matrix element is  related to the elastic scattering amplitude $f(s,b)$ by the known relation $S(s,b)=1+2if(s,b)$ and
can be written in the form
\begin{equation}
S(s,b)=\kappa(s,b)\exp[2i\delta(s,b)]
\end{equation}
with the two real functions $\kappa(s,b)$ and $\delta(s,b)$. 
The function $\kappa$ ($0\leq \kappa \leq 1$) is a transmission factor: its value $\kappa=0$ corresponds to a complete absorption of the initial state. 
At high   energies the real part of the scattering amplitude is small and can  be neglected, at least for
the purposes of the qualitative considerations, then  the substitution $f\to if$ with $f$ being a real function can be used.
 Therefore, the function $S(s,b)$ is also real, but  it could be positive or negative.

In fact, the choice of elastic scattering mode, namely, absorptive  or reflective one, is defined by the sign of the function $S(s,b)$, i.e. by the phase $\delta(s,b)$
\cite{ttprd}. The standard assumption is   $S(s,b)\to 0$  at the fixed impact parameter $b$ and $s\to \infty$. This  is called a black-disk limit, and  the elastic scattering  is completely absorptive.  In this case the function $S(s,b)$ is
always non-negative and it implies the limitation $f(s,b)  \leq 1/2$.  This absorptive approach has been used in the papers \cite{b1,b2,ros,fag}. To study the edge properties there were proposed to
consider the following difference: \begin{equation}
 f(s,b)-2f^2(s,b).
 \end{equation}
 It is equal to zero at $f=1/2$ ($S=0$, complete absorption) and at $f=0$ ($S=1$, no scattering).  Integration of this expression over impact parameter
$b$ results in the difference of the inelastic and  elastic cross--sections: 
\begin{equation}
\sigma_{tot}(s)-2\sigma_{el}(s)=\sigma_{inel}(s)-\sigma_{el}(s).
\end{equation}
 The existence of the edge with energy-independent width was postulated in \cite{b1} and confirmed on the base of the particular
eikonal model in \cite{b2}. These results have been obtained in the absorptive approach leading to the asymptotic limit 
\begin{equation}\label{din}
\sigma_{diff}(s)/\sigma_{inel}(s)\to 0
\end{equation} 
at $s\to\infty$. Eq. (\ref{din}) is at variance with the $t$-channel definition of the inelastic diffraction associating this process with the Pomeron exchange \cite{usdiff}.

There is  an alternative (to the absorptive approach) option: the function \[ S(s,b)\to -1\] at fixed $b$  and $s\to \infty$, i.e.  $\kappa \to 1$ and $\delta = \pi/2$. Such phase can be interpreted as the geometric phase related to the presence of singularity \cite{intja,arh13}.
General principles (unitarity) allow the function $S(s,b)$ to be negative in the certain region of $s$- and $b$ - values (i.e. at $s>s_0$ and $0\leq b<r(s)$, where $S(s,b=r(s))=0$) since the probability is a quadratic over $S$.
The function $S$ has negative values, in particular, in the Donnachie--Landshoff (DL) model (cf. \cite{dl} and the references therein) at the LHC energies. However, this model does not preserve unitarity and the value of  $|S(s,b)|$ eventually exceeds unity at fixed impact parameter when the energy is high enough. At the LHC energies the amplitude in this case, is exceeding the black-disk limit at small impact parameters, but, the amplitude and $S(s,b)$ are still lower than the corresponding unitarity limitations  (cf. \cite{adm}). 

As it was already noted, exceeding the  black-disk limit is a principal conclusion of the model--independent analysis of the impact parameter dependencies performed in \cite{alkin}.  This analysis has shown that
$f(s,b)$ becomes greater than the black-disk limit of $1/2$ at $\sqrt{s}=7$ TeV, but the relative positive deviation $\alpha$ ($f(s,b)=1/2[1+\alpha(s,b)]$) is still small at this energy. The value of $\alpha$ is about $0.08$  at $b=0$ \cite{alkin}. 
It should be stressed therefore, that the most relevant objects to study  deviations from the black-disk limit are the functions $f(s,b)$ and $h_{el}(s,b)$, but not the inelastic overlap function $h_{inel}(s,b)$ since relative deviation in the latter function is of order $\alpha^2$, namely $h_{inel}(s,b)=1/4[1-\alpha^2(s,b)]$, where $\alpha(s,b)$ is positive in the region $0\leq b<r(s)$.

The impact parameter scattering amplitude  at small $b$ values is sensitive to the $t$--dependence of the scattering amplitude $F(s,t)$ in the region of
large values of $-t$.  Therefore, it is not surprising that the models not reproducing positive deviation from black-disk limit, in particular, the eikonal models, provide a rather poor description of the LHC data in this region of transferred momenta \cite{b2,godiz,drufn}.  In contrast,
the DL model, in which  the black disk limit is exceeded, is in a good agreement with the new experimental data on $d\sigma/dt$ at $\sqrt{s}=7$ TeV in the whole region of transferred momentum\cite{dln}.

The limiting case  $ S(s,b)\to -1$ at fixed $b$ and $s\to \infty$ can be interpreted as a pure reflective scattering similar to reflection of light in optics \cite{intja}. 
The appearance of the reflective scattering can be associated with increasing density of a scatterer with the energy growth. It can be said that beyond the critical value of density, corresponding to the black-disk limit, the scatterer starts to reflect the initial wave like the metals do changing phase of the incoming wave by $180^0$. Thus, an increasing reflection appears in addition to a decreasing absorption. 
The principal point of the reflective scattering mode is that  $1/2  < f(s,b) \leq 1$ and $0  > S(s,b) \geq -1$, as allowed by unitarity relation \cite{ph1,phl}.

Now, one can  consider properties of the edge related to its energy dependence in the elastic scattering in the case when unitarity limit $1$ for the amplitude
$f(s,b)$ is supposed to be saturated instead of saturation of the black-disk limit $1/2$. Evidently, to this end one should consider the difference 
\begin{equation}
f(s,b)-f^2(s,b), 
\end{equation}
which is equal to zero at $f=1$ ($S=-1$, complete reflection) and at $f=0$ ($S=1$, absence of scattering). By analogy with consideration \cite{b1} it is to be referred as an edge. But,  according to unitarity relation, it is just  the inelastic overlap function $h_{inel}(s,b)$. The latter has a peripheral impact--parameter dependence  with maximum at $b=r(s)$ at high energies. The existence of such black ring has been discussed earlier (cf. e.g. \cite{intja}).

 The energy dependencies of $r(s)$ and $\sigma_{inel}(s)$ at $s\to \infty$ are similar,  i.e. 
 $r(s) \sim \ln s$ and $\sigma_{inel}(s) \sim \ln s$ (cf. e.g. \cite{usmpla}). The total cross--section of inelastic interactions is represented by the following integral over $b$:
\begin{equation}\label{inel}
\sigma_{inel}(s)=8\pi\int_0^{\infty} bdb h_{inel}(s,b),
\end{equation}
which can be estimated due to peripheral form of $h_{inel}(s,b)$ at high values of $s$ as 
\begin{equation}\label{ineles}
\sigma_{inel}(s)\simeq 8\pi r(s)\int_0^{\infty} db h_{inel}(s,b),
\end{equation}
Since both the inelastic cross-section and $r(s)$ increase at high energies logarithmically, one can
conclude on the energy independence of the integral 
\begin{equation}\label{i1}
I_{inel}(s)= \int_0^{\infty} db h_{inel}(s,b)
\end{equation}
at high energies.
The numerical value of $I_{inel}$ is a model--dependent one. Using the chiral quark model \cite{cqmus} one can obtain that
\begin{equation}\label{i3}
I_{inel}(s)\to \frac{ \xi}{ M_Q }
\end{equation}
at $s\to\infty$.  In Eq. (\ref{i3}) $M_Q$ is the total mass of constituent quarks in the two colliding hadrons
and the model dimensionless flavor-independent parameter $\xi$ has value close to 2. Thus, asymptotical value of $I_{inel}$ is about $0.2$ fm and it provides a reasonable estimate for the width of the edge.

Despite that the above sum rule and limiting behavior of $I_{inel}(s)$  is a specific feature related to the saturation of the unitarity bound on the partial amplitudes,  it is in agreement with the conclusion on   asymptotical energy independency of the edge in the case of the black-disk limit saturation \cite{b1} .

\section{The edge and  $d\sigma/dt$ at small-$t$ and high energies}
Unitarity for the elastic scattering amplitude $F(s,t)$ 
(we continue to consider pure imaginary case and use the substitute $F \to iF$) has the form
\begin{equation}\label{unit}
F(s,t)=H_{el}(s,t)+H_{inel}(s,t).
\end{equation}
In the case of unitarity saturation the approximate forms for the elastic and inelastic overlap functions are the following \cite{usmpla}:
\begin{equation}\label{hel}
H_{el}(s,t)\sim \frac{r(s)J_1(r(s)\sqrt{-t})}{\sqrt{-t}}
\end{equation}
and
\begin{equation}\label{hinel}
H_{inel}(s,t)\sim {r(s)J_0(r(s)\sqrt{-t})}.
\end{equation}
The inelastic overlap function $H_{inel}(s,t)$ plays a dual role and represents contribution of the edge into the elastic amplitude  in case of the unitarity saturation. Similarly, the elastic overlap function 
$H_{el}(s,t)$ is the contribution of the reflecting disk. Thus, in case of unitarity saturation, separation of the disk and edge contributions to the elastic amplitude is most naturally provided by the unitarity relation. 

 At small values of $-t$ the functions $J_{0,1}(x)$ can be approximated by the exponential functions
of different arguments. This means that $d\sigma/dt$ at small-$t$  can be approximated by three exponential functions with different arguments, too. It can, in principle, qualitatively explain the recent TOTEM result on the deviation of the dependence of $d\sigma/dt$ at small-$t$ from a simple linear exponent over $-t$ \cite{totem} due to the edge contribution. Quantitative analysis of these new data deserves separate description and will be reported elsewhere. We would like to note here  that the observed deviation of $d\sigma/dt$ behavior at the LHC from a simple  exponential one could be an argument for the peripheral form of $h_{inel}(s,b)$ resulted from the emergent reflective scattering mode at the LHC energies. In favor of this explanation is the fact that such deviation has not been observed at Tevatron and RHIC. At these energies the profile of  $h_{inel}(s,b)$ is central.\section{Interpretation of the reflective scattering}
The distinctive feature of the reflective scattering mode is a peripheral impact parameter distribution  
of the inelastic production probability. This fact in its turn can be associated with  production of the hollow fireball in the intermediate state of hadron--hadron interaction.
The projection of this fireball onto the transverse plane looks like a black ring.
The interpretation of this hollow fireball can be borrowed from the papers written about two decades ago \cite{bj,bj1}.
It was supposed  that the interior of this fireball is filled by the disoriented chiral condensate.  Its irradiation is supposed to be coherent, classical with a given isospin in each event, i.e. in one event the isospin can be 
directed along with $\pi^0$ while in other event its direction can be orthogonal leading to production of charged pions. 

Another possible interpretation of the reflective scattering mode can be made on the base of the already mentioned optical analogy.
Indeed, as it is well known the phase of incoming scattering wave is changed by $180^0$ at the scattering of light on metallic surface due to presence of  free electrons. Using this analogy, one can relate appearance of the reflective scattering mode with a transition to a deconfined phase of matter under hadron collisions above some threshold energy.

\section*{Conclusion}
We have discussed here the geometrical properties of edge in the case of unitarity saturation.  In this asymptotical scattering picture the interaction region can be represented by the reflecting disk with the black edge which is due to the inelastic interactions in this case. Decomposition of the elastic amplitude into  central reflecting disk and peripheral edge contributions is naturally provided by the unitarity relation in the considered case.
\section*{Acknowledgements}
We are grateful to D. Fagundes, L. Jenkovszky, V. Petrov, and P. Silva for the interesting discussions .
One of us (S.T.) is also grateful to R. Schicker and A. Szczurek, the organizers of  the school on ``Diffractive and electromagnetic processes at high energies'' held in Bad Honnef, Germany, where this note has been finished, for the warm hospitality. 


\small
\begin{thebibliography}{99}
\bibitem{y1}
T.T. Chou, C.N. Yang, Phys. Rev. {\bf 170} (1968) 1591.
\bibitem{y2}
T.T. Chou, C.N. Yang, Phys. Rev. Lett. {\bf 20} (1968) 1213.
\bibitem{y3}
T.T. Chou, C.N. Yang, Phys. Rev. {\bf 175} (1968) 1832.
\bibitem{b1}
M. M. Block, L. Durand, F. Halzen, L. Stodolsky, and T. Weiler, Phys. Rev. D {\bf 91}, (2015) 011501(R). 
\bibitem{b2}
M. M. Block, L. Durand, Phuoc Ha, and F. Halzen,
arXiv:1505.04842 [hep-ph].
\bibitem{ros}
J. Rosner, Phys. Rev. D {\bf 90},  (2014) 117902. 
\bibitem{fag}
D. A. Fagundes, A. Grau, G. Pancheri, Y. N. Srivastava, and O. Shekhovtsova, Phys. Rev. D {\bf 91}, (2015) 114011. 
\bibitem{valin}
R. Henzi, P. Valin,
Phys.Lett. B {\bf 160} (1985) 167.
\bibitem{valin1}
R. Henzi, P. Valin,
Phys.Lett. B {\bf 164} (1985) 411. 
\bibitem{ph1}
S.M. Troshin, N.E. Tyurin, Phys. Lett. B {\bf 208}  (1988) 517.
\bibitem{phl}
S.M. Troshin, N.E. Tyurin, Phys. Lett. B {\bf 316}  (1993) 175.
\bibitem{degr}
P. Desgrolard, L.L. Jenkovszky, B.V. Struminsky, 
Phys. Atom. Nucl. {\bf 63} (2000) 891.
\bibitem{intja}
S.M. Troshin, N.E. Tyurin, Int. J. Mod. Phys. A {\bf 22}  (2007) 4437.
\bibitem{alkin}
A. Alkin, E. Martynov, O. Kovalenko, and S.M. Troshin, { Phys. Rev. D} {\bf 89} (2014) 091501(R).
\bibitem{dremin}
I.M. Dremin,  Adv. High Energy Phys. {\bf 2015} (2015) 912743.
\bibitem{8tev}
G. Antchev et al. (The TOTEM Collaboration)
Phys. Rev. Lett. {\bf 111} (2013) 012001.
\bibitem{ttprd}
S.M. Troshin, N.E. Tyurin, Phys. Rev. D {\bf 88}  (2013) 077502.
\bibitem{usdiff}
S.M. Troshin, N.E. Tyurin, arXiv:1503.03612 [hep-ph].
\bibitem{arh13}
S.M. Troshin, N.E. Tyurin,  arXiv:1305.6153 [hep-ph].
\bibitem{dl}
A. Donnachie, P.V. Landshoff, Phys. Lett. B {\bf 296}  (1992) 227.
\bibitem{adm}
A.D. Martin, Lecture given at { \it Diffractive and Electromagnetic Processes at High Energies}, Acquafredda, Italy, September 6-10, 2010.
\bibitem{godiz}
A.A. Godizov,
Eur. Phys. J. C{\bf 75} (2015) 224.
\bibitem{drufn}
I.M. Dremin, Physics-Uspekhi, {\bf 56} (2013) 3.
\bibitem{dln}
A. Donnachie, P.V. Landshoff, 
Phys. Lett. B{\bf 727} (2013) 500.
 \bibitem{usmpla}
S.M. Troshin, N.E. Tyurin, Mod. Phys. Lett. A {\bf 24} (2009) 1103.
\bibitem{cqmus}
S.M. Troshin, N.E. Tyurin, Phys. Rev. D {\bf 49}  (1994) 4427.
\bibitem{totem}
G. Antchev et al. (The TOTEM Collaboration),  arXiv:1503.08111 [hep-ex].
\bibitem{bj}
J.D. Bjorken, K.L. Kowalski, C.C. Taylor,
SLAC-PUB-6109, 1993.
\bibitem{bj1}
G. Amelino-Camelia, J. D. Bjorken, S. E. Larsson,
Phys. Rev. D {\bf 56}, 6942, (1997).
\end{thebibliography}
\end{document}